\documentclass[twocolumn,showpacs,preprintnumbers,amsmath,amssymb]{revtex4}

\usepackage{graphicx}
\usepackage{dcolumn}
\usepackage{bm}

\begin{document}
\preprint{APS}
\title{Short Necklace States, Logarithm Transmission Fluctuation and Localization Length}
\author{Xunya Jiang}
\email{xyjiang@mail.sim.ac.cn}
\author{Liang Chen}
\affiliation{State Key Laboratory of Functional Materials for Informatics,\\
Shanghai Institute of Microsystem and Information Technology, CAS, Shanghai 200050, China\\
Graduate School of the Chinese Academy of Sciences, China}


\begin{abstract}
We investigate the widely-existing short necklace states in random
systems. It is found that their peak width and relative height in
$lnT$ spectra keep almost constant when the system length increases,
which is explained by the coupled-resonator theory with intrinsic
parameters. This property makes them special in contribution of
$lnT$ fluctuation. Further, short necklace states can help us to
deeply understand the physical meaning of localization length and
the delocalized effect in localized regime.
\end{abstract}


\maketitle


Anderson localization\cite{AndersonLocalization} has changed the
basic propagating picture of waves(classical and quantum) in random
systems and is used to explain phenomena, \emph{e.g.}
metal-insulator transmission and quantum Hall effects etc. The
pioneering studies of quasi one-dimensional (1D)
systems\cite{AdditivelnT} reveal that the logarithm transmission
$lnT$, not the transmission $T$, is statistically
Gaussian-distributed and the mean value $\langle lnT \rangle$ over
many configurations is linearly additive with the system length $L$.
Based on this, the localization length which is the most essential
length scale in localization study can be naturally defined as
$\xi=-2L/\langle lnT \rangle$. However, unlike $T$ whose fluctuation
has been widely studied, surprisingly the $lnT$ fluctuation has not
been intensively studied to the best of our knowledge. Obviously, if
the origin of the $lnT$ fluctuation is well understood, the physical
meaning of localization length can be more deeply revealed.

On the other hand, not all states are localized even in the strongly
localized regime, as pointed out in
Ref.\cite{Pendry1987,Tartakovskii} that the degenerated localized
states can form ``necklace states" (NSs) through a 1D random
configuration. NSs can generate the \emph{mini-bands}, \emph{i.e.}
in transmission spectra, which dominate the ensemble average of
conductance \cite{Pendry1987,Pendry1994}. After the NSs are observed
in random optical 1D systems \cite{Bertolotti2005PRL,Sebbah2006PRL},
some studies have been done, such as the NS probability
\cite{Bertolotti2006PRE} and transmission
oscillation\cite{Ghulinyan2007PRL,Ghulinyan2007PRA} because of
different optimal order of NS. Our previous work \cite{Chen2011NJP}
also demonstrates that NSs do not follow the
single-parameter-scaling theorem. Recently, Ping Sheng propose that
NSs may take a critical role in Anderson phase transition(APT) in
his review on Science\cite{ShengPingPercolation}. Such proposing
shows that many details of APT are still not well understood and the
relation between APT and NSs needs more study. But numerically and
experimentally, for $L \gg \xi$ configurations, it is found that NSs
are so extremely rare that it is hard to find one even in millions
of configurations. Then it is rational to doubt whether NSs are
qualified to take the critical role in APT or important for
localization study. If some kinds of NS widely exist in \emph{almost
all} random configurations and almost all frequency ranges, then the
role of NS could be changed radically.

After careful review, we find that previous NS studies are focused
on \emph{long necklace state}(LNS) which is spatially extended
through a whole configuration. The length is an important property
of NS, but it is secondary. The most essential property of NS is the
coupling between localized states which are near to each other in
\emph{both} frequency domain (nearly degenerated) and in spatial
domain so that they form a ``chain"(necklace) and help wave to hop
from one to another. If we disregard the length requirement,
\emph{e.g.}, a few localized states form a chain which is so short
that it is deeply embedded inside a long configuration, we define
such a chain as \emph{short necklace state}(SNS). Obviously, the
occurrence probability of SNS is much much higher than LNS. Then the
question becomes ``Are there special properties of SNSs which make
them very important in localization studies?"

In this Letter, we will demonstrate that the high plateau in $lnT$
spectra are the evidence of widely existing SNSs. SNSs are an
important origin of the $lnT$ fluctuation and very essential for
fully understanding the physical meaning of localization length. The
most special property of SNSs is that their peak width and the
relative hight in $lnT$ spectra depend only on intrinsic parameters,
i.e. the coupling strength between localized states and the length
of SNS, and is independent of configuration length $L$. This
property is found numerically and also derived by our theory based
on the coupling-resonator model. So SNSs have priorities over LNSs
since their much higher occurrence probability, and also have
priorities over localized states since their \emph{ much wider and
almost $L$-independent} peak-width. Further more, we demonstrate
that the localization length $\xi$ is not a simple `decay length'
since it includes the delocalized effects of SNSs. With all these
results, a basic physical picture can be formed in which SNSs are
the widely existing seeds of delocalization effect even in the
strong localized regime and can help us to understand details of
APT.

Our numerical 1D model is made of optical binary layers, \emph{i.e.}
each cell contains two kind layers, whose dielectric constants are
$\epsilon_1=1$ and $\epsilon_2=2$, and thicknesses are $d_1=95 nm$
and $d_2=d_2^0(1+W\gamma)$ where the randomness is introduced,
$d_2^0=120 nm$ is the average thickness, $\gamma$ is the random
number in the range of $[-1, 1]$ and $W$ is the randomness strength
which is chosen as 0.1 in this work.

Following the history of localization study, we start from the
statistic property of $lnT$ of configurations. The $lnT$
distribution of $2\times10^5$ configurations for 3000-cell and
6000-cell systems at frequency $f_0=2.5422\times10^{15}$Hz are shown
in Fig.1. Same as expected, $lnT$ is Gaussian distributed. From the
$\langle lnT \rangle$, we can obtain $\xi = -L/ \langle lnT \rangle
= 150 a_0$, where $a_0=d_1+d_2^0$ is the average cell length and
chosen as the length unit. The Gaussian distribution of $lnT$ is
well known, but \emph{the physical reasons of $lnT$ fluctuation},
which related with basic understanding of localization length, is
still waiting for more detailed study.

\begin{figure}
\includegraphics[width=0.8\columnwidth]{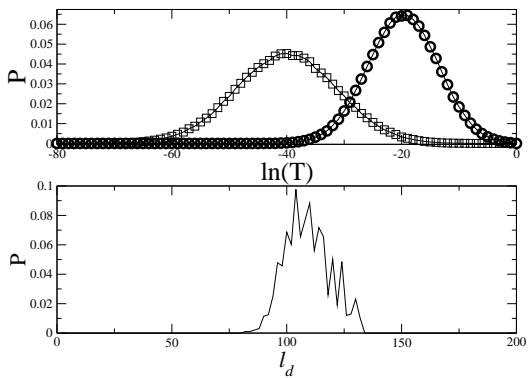}
\caption{Fig.1a, the $lnT$ distribution of 3000-cell and 6000-cell
configurations. Fig.2b, the distribution of decay length $l_d$ of
localized states in $L=10\xi$ configurations.} \label{PlnTPld}
\end{figure}
The first naive thought of the $lnT$ fluctuation is that it's
originated from the resonant transmission peaks of localized states,
such as Azbel states\cite{Azbel}. But from simple scaling argument,
we will find out these peaks are negligible in $lnT$ fluctuation for
enough long systems, $e.g.$, $L > 10 \xi$. The contribution of
localized state peaks can be roughly estimated as $P*H$, where $P$
is the probability of a frequency exactly falling on a resonant
peak, and $H \sim L/\xi$ is the averaged hight of peaks in $lnT$
spectra. $P$ is $\propto \rho {\delta \omega}$, where $\rho \propto
L$ is the density of state and ${\delta \omega}$ is the peak average
width of localized states. Since peakwidth $\delta \omega \propto
e^{-L/\xi}$ decrease exponentially with L, the contribution to $lnT$
fluctuation from localized state peaks can be neglected when $L \gg
\xi$.

The second naive thought is that the $lnT$ fluctuation is from the
different decay length $l_d$ of the localized states. But after
numerical calculation of $l_d$ from field distribution of localized
states in $3\times10^4$ random configurations with $L=10\xi$, we
find that the distribution of $l_d$, shown in Fig. 1b, is quite
different from that of $lnT$. We can see that the distribution of
$l_d$ is not Gaussian and its mean value is about $l_d \sim 100$
much smaller than $\xi=150$. The difference between $l_d$ and $\xi$
will be discussed later.

\begin{figure}
\includegraphics[width=0.8\columnwidth]{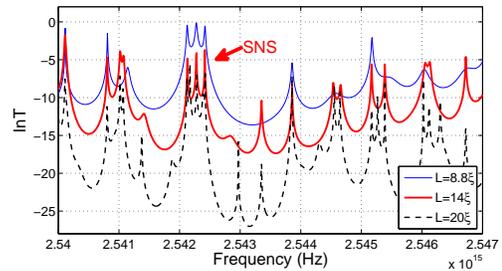}
\caption{The spectra of three configurations. The longer one is
constructed by adding layers to shorter ones.} \label{lnTSpectra}
\end{figure}
To find the physical reason of the fluctuation of $lnT$, we need to
check the $lnT$ spectra of random configurations for details. Three
typical $lnT$ spectra are shown in Fig.2 of random configurations
with length $8.8\xi$, $L=14\xi$ and $L=20\xi$. The longer
configuration in Fig.2 is constructed by adding more layers to the
shorter configuration(keeping the original part unchanged). At first
sought, the spectra are like the ``fish backbone" with the sharp
resonant peaks of localized states as ``stings". But we note that
the base of backbone is not quite flat and there are plateaus and
valleys which are higher or lower than the mean value $\langle lnT
\rangle = - L / \xi$. These plateaus and valleys are the basic
spectral feature of one certain configuration. For \emph{totally
different} configurations, the feature (the positions of plateaus
and valleys) are totally different. Hence, it is rational to think
that the fluctuation of $lnT$ at certain frequency is mainly from
these plateaus and valleys of different configurations. Actually,
not only fluctuation of $lnT$, the mean value $\langle lnT \rangle$,
which defined the localization length $\xi$, is also averaged over
all valleys and plateaus for different configurations at certain
frequency. Then the question becomes ``where are these plateaus and
valleys from?".

\begin{figure}
\includegraphics[width=0.8\columnwidth]{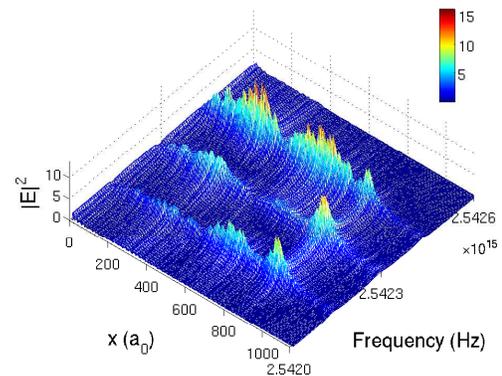}
\caption{The field intensity $|E|^2$ versus $x$ and $f$ of the
necklace state signed in Fig.2} \label{FieldSpectra}
\end{figure}
After careful observation, we find that, although the plateaus and
valleys are quite different for totally different configurations,
\emph{the spectral feature has some kind of inheritance, if we
increase the system length $L$ by adding more layers at both sides
of the original configuration, as we did in Fig.2}. The inheritance
of the spectra is clearly shown in Fig.2, where the longer
configuration is constructed by adding more layers to the shorter
one. The inheritance infer that these plateaus and valleys are not
sensitive to the boundary of a configuration, or in other words,
they are really from intrinsic things deeply inside a configuration.
After careful check, we find that \emph{the origin of the plateaus
in transmission spectra is the SNS which widely exist deeply inside
long configurations}. From Fig.2, we can clearly see that, for short
configuration with $L=8.8\xi$, there is a LNS signed by red arrow
which is same as that predicted by Pendry and Tartakovskii
\cite{Pendry1987,Tartakovskii}. The field distribution at resonant
frequencies shown in Fig.3 also confirm the judgement. After adding
more layers, for $L=24$ configuration, the plateau is still there,
but much lower and now the LNS becomes SNS which deeply embedded
inside the longer configuration. Inversely, if we find a SNS plateau
in the spectra of a long configuration, we can locate the position
of SNS by the field distribution at the resonant frequencies, then,
by removing more layers far away from the SNS, we change the SNS
into LNS of a shortened configuration. So, SNS and LNS could be
transformed into each other by adding or removing more layers. But
for very long configurations, compared with extremely rare LNS, SNS
widely exist in almost all frequency range.

With SNS concept in random configurations, we can understand the
$lnT$ spectra of random configurations in a very different picture
now. Without SNS, the spectra of random configurations are formed by
randomly distributed peaks of independent localized states, so that
the fluctuation is only from fluctuation of density of states. But
now, we know that the coupling between a few localized states can
generate correlation in $lnT$ spectra, which is represented by many
small plateaus. Actually, compared with LNS which generate
correlation in $T$ spectra which have been studied by Pendry and our
previous work \cite{Chen2011NJP}, SNS can generate correlation in
$lnT$ spectra \cite{chenjiang}. Widely existing SNS plateaus form
the basic feature of $lnT$ spectra, as shown in Fig.2. But SNS
population is still much less than localized states. Next, we will
show that an important property of SNSs makes them much more
important than localized states as the source of $lnT$ fluctuation.

\emph{Comparing the signed plateaus in Fig.2 with different $L$, we
find that their width and relative hight keep almost constant,
although the absolute hight is reduced with $L$}. This property
makes SNS qualitatively different. The contribution of localized
states to the $lnT$ fluctuation can be neglected for long systems
($L \gg \xi$) since their peak-width decreases exponentially with
length, while the SNS contribution of $lnT$ spectra is much larger
since their constant peak width and relative hight.

Why the width and hight of SNS are almost constant for different
$L$? And, with increasing length L by adding more layers, why most
plateaus are robust but some of them will split? We will reveal the
physics behind these SNS properties. From different behaviors of
plateaus, we can separate the SNS plateaus into two classes, the
\emph{short intrinsic NS} (SINS) for the robust ones and \emph{short
non-intrinsic NS} (SNNS) for the splitting ones. We use the
coupled-resonator model to study the difference between two classes.
Suppose that two localized states, whose frequencies and positions
are near to each other, can be represented by two coupled
resonators:
\begin{eqnarray}
 \ddot{E_1} + \delta_1 \dot{E_1} + \omega_1^2 E_1 = \kappa_{12}
\dot{E_2} \nonumber
\\
 \ddot{E_2} + \delta_2 \dot{E_2} + \omega_2^2 E_2 = \kappa_{21}
\dot{E_1}
\end{eqnarray}
where $E_i$, $i=1$ or $2$, are the fields of two localized state,
and $\dot{}$ means the time derivative; $\omega_i$ and $\delta_i$
are the ``original" central frequencies and peak-widths (inverse of
quality factors) of two localized states if without coupling, and
$\kappa_{12}=-\kappa_{21} \propto \int dx (E_1* E_2)$ are the
coupling strength between two localized states which proportional to
the overlap integral. To classify these different behaviors, we need
to compare scales of several parameters in Eq.(1).
 First, supposing that the sum of original peak-widths $\delta_1+\delta_2$ is
larger than peak distance $\Delta \omega _{12} = | \omega_1 -
\omega_2 |$ which means two peaks overlapping with each other. For
example, a configuration length $L$ is not very long so that two
peaks are wide enough to overlap with each other, then, two peaks
form a plateau. But, when the length $L$ increases by adding more
layers, both peak widths $\delta_i$ decrease exponentially, then,
when $\delta_1+\delta_2 < \Delta \omega _{12} $, whether the plateau
will split or not depends on a new condition, the comparing between
$|\kappa_{12}|$ and $|\Delta \omega_{12}|$. From the
coupled-resonators theory \cite{Landau}, we can obtain that,
\emph{if the condition}
\begin{eqnarray}
\kappa_{ij}
> \Delta \omega = \omega_i - \omega_j
\end{eqnarray}
 \emph{is satisfied, two
resonators are always well-coupled to each other and now the plateau
width is determined by $\kappa_{ij}$(peak-repulsion effect),
independent of any other conditions }. If this condition is not
satisfied, the plateau will split when $L$ is large enough. For
multi-coupled localized states of $n$th-order SNS, the discussion is
similar. Now, the reason of different behaviors of plateaus is
obvious, for the SINS, the plateau is very robust and its width is
determined by coupling strength, but for SNNS, the plateau will
split. For long $L>>\xi$ configurations, most plateaus are from the
SINS which are deeply embedded inside the configurations, while very
few exceptions are generally near the configuration edges since the
peak width of near-edge localized states is much larger. In
following discussion of this work, all SNS are referred to SINS.
From Eq.(2), we can also obtain the optimal distance of localized
states in SNS \cite{optd}.

\begin{figure}
\includegraphics[width=0.8\columnwidth]{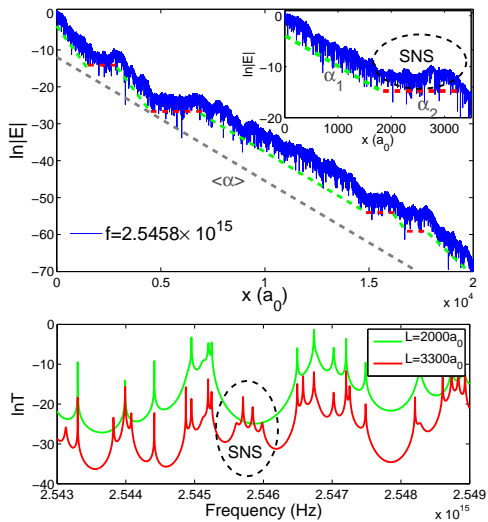}
\caption{Fig4a, the $ln|E(x)|$ vs $x$ of a configuration with
$2\times10^4$ cells, and the ``slope sections" are singed by dashed
lines. The insert is the enlarged part of first $3000$ cells. Fig4b,
the $lnT$ spectra of two configurations with first $2000$ cells
(first slope section) and $3300$ cells(including first and second
slope section) of Fig.4a configuration.} \label{lnE}
\end{figure}
Next, we will demonstrate that the localization length $\xi=-L/
\langle lnT \rangle $, the most important length scale in Anderson
localization theory, is related with SNS too. It's well known that
there is a ``save" way to obtain the precise $\xi$ numerically,
.i.e. calculating the electric field $E(x)$ of a $very$ $long$
random configuration at a chosen frequency, then $\xi$ can be obtain
as the inverse of slope $\xi \simeq -L/ln(E(L)/E(0))$. The violation
will be very small for very large $L$, since the variance $\sigma
\propto \sqrt{L / \xi}$ is much smaller than mean value $-L/\xi$,
which is called as self-averaging property. However, if we are
careful, more interesting details can be found as shown in Fig. 4a,
where $ln |E|$ vs $x$ at frequency $f=1.5996 \times 10^{16}$Hz is
shown. The insert is the enlarged first small part of Fig.4a for
detailed discussion. We find that, except short-range fluctuations,
there are several long-range ``sections" signed by dashed lines as
shown in Fig.4a. In each section, the slope $\alpha_i$ does not have
large change, but in different sections the slopes could be quite
different, which is clearly shown in $\alpha_1$ and $\alpha_2$ in
Fig.4a insert. Where are these "sections" from? We can find the
reason if we check two $lnT$ spectra shown in Fig.4b, which are from
two configurations with the $2000$ cells and the first $3300$ cells
of the long configuration in Fig.4a. The first configuration
($I_{conf}$) exactly includes first section of Fig.4a insert, while
the second configuration ($II_{conf}$) includes both first and
second sections. For $I_{conf}$ whose slope in Fig.4a is very
larger(compared the average slope), we find that the frequency is
exactly falling into the spectrum valley. But for $II_{conf}$ which
ends with very-small-slope section, we find that there is a SNS
appearing at the frequency, and again the inheritance is very clear
for two spectra. Obviously, the occurrence of SNS has strong
influence on the local slope and determined a ``very-small-slope
section". Another important influence of slope is from fluctuation
of density of states(very few states in the valley), but it is not
the main topic of this work. With these observation, we can
construct such a more detailed picture to describe the decay slope
of $ln|E|$ when configuration length $L$ increases: if a SNS occurs,
the slope will be much smaller, but if without SNS and even very few
localized states occurs, the slope will be much larger. Based on SNS
picture, three conclusions are derived from observation. The first
is that the typical ``section" length (about $10\xi$ in our model)
actually is the typical length scale $l_{NS}$ for a SNS to appear at
certain frequency. Second, we have new understanding of the
localization length $\xi$, since, for a very long configuration,
many short SNS in it have strong influence not only on the slope of
sections, but also on the total average slope $\overline{\alpha}$,
which is the inverse of $\xi$. Third, we have new understanding of
self-averaging property of $lnT$ spectra too, \emph{i.e.} the
average slope of a very long system has not only averaged the
density fluctuation of localized states, but also averaged the
occurrence fluctuation of SNS. Our recent work \cite{Chen2011NJP}
shows that the occurrence probability $P_{NS}$ of LNS will increases
considerably near APT point, so that NS is not confined by single
parameter theorem and could take a critical delocalization role at
APT. In this work, we can see that, even in very strong localized
regime, SNS represent widely existing delocalization mechanism. SNS
may can tell us more details of the APT scenario, \emph{i.g.} more
and more SNS can be connected with each other and form a LNS at
last, so that the whole configuration is conducting.

At last, we hope to discuss the decay length $l_d$ of localized
states and its relation with $\xi$. As we have shown in Fig.1b, the
$l_d$ typical value of configurations with $L=10\xi$ is very
different from $\xi$, which is against our common sense since we
generally use $exp(-L/\xi)$ to represent the field of localized
states. Numerical results tell us that, for not-very-long $ L < 2
l_{NS} = 20\xi$ configurations, $l_d$ typical value is smaller than
$\xi$. Our explanation is following: since there is no SNS in such
not-very-long system generally, $l_d$ is the ``naked" decay length
which is corresponding the large slope in Fig.4a, and the field
should be write as $exp(-L/l_d)$. Only in the very long range ($x
>> l_{NS}$),  the localized field can approximated as $exp(-|x|/\xi)$
since all effects of SNS are included.

In summary, the SNSs which widely exist in long random
configurations are investigated. The speciality of SNSs and their
contribution to $lnT$ fluctuation and localization length $\xi$ are
studied numerically and theoretically. With SNSs, a much more
detailed picture of Anderson phase transition can be set up, and
many topics can be quantitatively studied, such as the evolution of
SNS with strength of randomness, the change of localization length
and its relation with evolution of SNS,  the correlation in $lnT$
spectra and the optimal order of SNS. These studies will reveal more
details of APT in future.


\begin{thebibliography}{s1}
\bibitem{chenjiang} The studies of $lnT$ correlations will be published elsewhere.
\bibitem{optd} We can evaluate the ``optimal distance" of localizes states in SNS by the $\partial P_d / \partial d |_{d=d_{opt}} = 0 $,
where $P_d$ is the probability for a localized state to find another
nearly-degenrated localized state in distance $d$.  Since $P_d
\propto d \Delta \omega \sim d \kappa_{ij} $ and $\kappa_{ij}
\propto exp(-d/l_d)$ where $l_d$ is the decay length of localized
field, we can obtain the optmal distance as $d_{opt}=l_d$, which
agree with our numerical observation very well.
\end{thebibliography}
\end{document}